\documentstyle[epsf,a4,12pt]{article}
\begin{document}

\rightline {Si-97-01 \  \  \  \   }
\vspace*{2.truecm} 
\parskip 8pt 
\begin{center}
{\Large \bf The short-time behaviour of a kinetic } 

\vspace{0.3cm}

{\Large \bf Ashkin-Teller model on the critical line}
\vspace{3.5ex}
\end{center}
\vspace{6.4ex}
\centerline{\sf Z.B. Li, X.W. Liu}
\vspace{2.5ex}
\centerline{\sf The Centre for 
Computational Physics, Zhongshan University, 
Guangzhou, China}
\vspace{2.5ex}
\centerline{\sf L. Sch\"ulke, B. Zheng}
\vspace{2.5ex}
\centerline{Universit\"at-GH Siegen, D-57068 Siegen, Germany}

\vspace{6ex}
\centerline{\bf Abstract}
\begin{center}
\vspace{2.5ex}
\begin{minipage}{5in}

{We simulate the 
kinetic Ashkin-Teller model with both ordered and disordered initial
states,
evolving in contact with a heat-bath at the critical temperature. 
The power law scaling behaviour for the magnetic order and electric
order
are observed in the early time stage.
The values of the critical exponent $\theta$ vary along
the critical line. Another dynamical exponent $z$ is also obtained
in the process.}
\end{minipage}
\end{center}
\vspace{0.5cm}

\vspace{0.3cm}


\vfill
\eject
It has been 
predicted by 
analytical calculations \cite {jan89}  and
supported in Monte Carlo simulations 
\cite {hus89,hum91,li94,men94,sch95} 
that there exists scaling in the macroscopic
short-time regime for some critical dynamic processes. The
universality
was confirmed for the Ising model and the Potts model with various
dynamics and
different lattice structures \cite {men94,oka97a,liu95}.
Basing upon the universal scaling hypothesis for the initial stage, 
promising methods have been proposed for numerical measurements of
critical exponents, including both static and dynamic ones
\cite {li95,li96,sch96}.
It has been
suggested that the critical point can be also determined in this stage
\cite {sch96}. 
The universal behaviour of the short-time dynamics is found to be
quite
general \cite {jan92,oer93,oer94,die93},
 e.g., in connection with ordering dynamics or
damage
spreading \cite {bra94,gra95} and surface critical phenomena 
\cite {rit95}. 

It is interesting to see the short-time behaviour of models
possessing continuously varying critical exponents 
in equilibrium. It was found by Kadanoff and Wegner that there is a
connection
between the continuous variation of the critical exponents and the
existence
of a marginal operator \cite {kad71}.
The variation of critical exponents in the early time evolution
implies that 
the operator keeps in marginal even in the non-equilibrium 
initial stage.

The Ashkin-Teller model is one of such models that has week
universality,
i.e., the critical exponents vary with parameters of the interaction
\cite {bax82}. It contains two species of spins $\{\sigma_{i}=\pm 1\}$ and
$\{\tau_{i}=\pm 1\}$, located on a square lattice. The interaction is
described by the Hamiltonian
\begin{equation}
H=\sum_{<i,j>}[K (\sigma_{i}\sigma_{j}+\tau_{i}\tau_{j})+
K_{4} \sigma_{i}\sigma_{j}\tau_{i}\tau_{j})]
\label{e1}
\end{equation}
This model is dual to the solved Baxter's eight vertex model. It has
five
phases. We only investigate the exactly known segment of the critical
line
that separates the disordered phase and the phase where both the 
magnetic order and electric order are not zero. Along that line,
the critical exponents vary continuously. The critical
line is given by the equation
\begin{equation}
\exp(-2K_{4})=\sinh(2K)
\label{e2}
\end{equation}
where $(\ln3)/4< K < +\infty $. Besides the critical exponent $\nu$
that
governs the critical behaviour of the correlation length,
there are other two independent critical exponents 
$\beta_{m}$ and $\beta_{e}$ corresponding to two order parameters, 
the magnetic order $<\sigma_{i}>$, and the electric order
$<\sigma_{i}\tau_{i}>$, respectively.
The critical exponents for the equilibrium state have been 
known as \cite {bax82,kad79a,kad79,kno80}
\begin{equation}
\nu=\frac{2-y}{3-2y}, ~~~~\beta_{m}=\frac{2-y}{24-16y},
~~~~\beta_{e}=\frac{1}{12-8y}
\label{e3}
\end{equation}
where the parameter $y$, ranging from $0$ to $4/3$, is related to
$K_{4}$
by the equation
\begin{equation}
\cos(\pi y/2)={1 \over 2}[\exp(4K_{4})-1]
\label{e4}
\end{equation}

When dynamics is introduced into the Ashkin-Teller model, besides 
the well-known $z$ that describes the 
divergence behaviour of the time-correlation length, 
there are two critical exponents related to the dimensions of 
two initial orders. 
As argued by Janssen et al., \cite {jan89} the initial scaling emerges at
fixed point
under the renormalizational group transformation. By naive coarsing 
transformation, one can see two fixed points for initial states, i.e.,
$m_{0}=e_{0}=0$ at very high temperature and $m_{0}=e_{0}=1$. Our
simulations
are around these two fixed points.

The magnetic moment $M^{(k)}$ and the electric moment $E^{(k)}$ are
defined
as
\begin{equation}
M^{(k)}=<[(\sum_{i}\sigma_{i})/L^{2}]^{k}>_{t},~~~~~~~~~~~~~~~~~~~
E^{(k)}=<[(\sum_{i}\sigma_{i}\tau_{i})/L^{2}]^{k}>_{t}
\label{e5}
\end{equation}
In analogy to the Ising
model, we suppose the following finite-size scaling relations hold 
in the vicinity of the first fixed point,
\begin{equation}
M^{(k)}(t,L,m_{0})=b^{-k\beta_{m}/\nu}M^{(k)}(b^{-z_{m}}t,b^{-1}L,b^{x_{m}}
m_{0})
\label{e6}
\end{equation}
for $e_{0}=0$, and $m_{0}$ small, and
\begin{equation}
E^{(k)}(t,L,e_{0})=b^{-k\beta_{e}/\nu}E^{(k)}(b^{-z_{e}}t,b^{-1}L,b^{x_{e}}
e_{0})
\label{e7}
\end{equation}
for $m_{0}=0$, and $e_{0}$ small. We have denoted the anomal
dimensions of
the initial magnetization and electric order as $x_{m}$ and $x_{e}$ 
respectively. For $k=1$, taking $b=t^{1/z}$ and assuming
the initial orders are small enough, we can
expand the magnetization(electric order) with respect to
$m_{0}(e_{0})$ and 
obtain
\begin{equation}
M(t)=m_{0}t^{\theta_{m}}F_{m}(t^{-1/z_{m}}L)
\label{e8}
\end{equation}
\begin{equation}
E(t)=e_{0}t^{\theta_{e}}F_{e}(t^{-1/z_{e}}L)
\label{e9}
\end{equation}
where $\theta_{m}=(x_{m}-\beta_{m}/\nu)/z_{m}$, and 
$\theta_{e}=(x_{e}-\beta_{e}/\nu)/z_{e}$. The scaling functions
$F_{m}$ and $F_{e}$ take account of the deviation from power laws due
to
the finite-size effect. They tend to constants as $L$ tends to
infinity and 
time tends to zero, but bigger than the microscopic time scale
$t_{mic}$.
It has been shown in various models that $t_{mic}$ is
ignorable in the heat-bath dynamic process. 
In the following fits, we will consider $F_{m}$ and $F_{e}$ as 
constants.

It has been stressed in Ref. \cite {jan89} that 
the initial 
states must have very short correlation lengths
and that a sharp 
preparation of the initial state improves the result.
For the Ashkin-Teller model, there are two initial order parameters,
$m_{0}$ and $e_{0}$. So one has more than one way to approach the 
disordered
fixed point. For instance, we can either let $e_{0}=0$ and $m_{0}$
small or
reverse, $m_{0}=0$ and $e_{0}$ small, later on referred to as 
initial condition I and 
initial condition II, respectively.
We find that the relaxation patterns from
these two initial conditions are very different. 
However, for both initial conditions, power laws are observed. The 
exponents $\theta$ depend on the initial conditions and vary with
$y$. 

\begin{table}[t]\centering 
$$
\begin{array}{|c|c|c|c|c|c|c|} 
\hline
  ~     &    L=180 &  \multicolumn{4}{|c|} { L= 60}    &
             L= 30      \\ \hline        
  y     &   m_{0}= 0.02  &   m_{0}= 0.02  &  m_{0}= 0.04   &
            m_{0}= 0.06  &   m_{0}= 0.08  &  m_{0}= 0.04   \\ \hline
  0     &   -.020(1) &   -.021(4) &   -.022(5) &   -.019(5) &  
  
            -.021(2) &   -.022(6)    \\ \hline
  1/6   &   -.013(1) &   -.014(2) &   -.014(5) &   -.013(4) &  
            
            -.014(2) &   -.015(6)    \\ \hline
  1/3   &    .009(2) &    .008(1) &    .007(5) &    .007(3) &   
             
             .007(3) &    .007(4)    \\ \hline
  1/2   &    .041(5) &    .041(3) &    .041(5) &    .041(2) &   
             
             .041(2) &    .039(6)    \\ \hline
  2/3   &    .085(4) &    .085(4) &    .085(5) &    .084(3) &    
             
             .083(2) &    .086(6)    \\ \hline
  5/6   &    .137(2) &    .137(2) &    .134(3) &    .136(3) &    
             
             .134(1) &    .135(3)    \\ \hline
  1     &    .191(2) &    .189(2) &    .188(3) &    .188(2) &   
             
             .185(1) &    .187(3)    \\ \hline
  7/6   &    .230(1) &    .229(2) &    .227(1) &    .225(1) &   
             
             .223(1) &    .225(3)    \\ \hline
\end{array}
$$
\caption{
Results for $\theta_{m}$ with initial condition I 
of $m_{0}=0.02$ on $L=180$, and of $m_{0}=0.02$, $0.04$, $0.06$, 
and $0.08$ on $L=60$, $m_{0}=0.04$ on $L=30$. 
In each case, the magnetization
in the time interval $1\le t \le 5$ is used for the least-square fit
to 
the power law.
}
\label{t1}
\end{table}

The short-time behaviour of the magnetization 
with the initial condition I can be seen in Fig.~1. 
The points are magnetizations
averaged over $40000$ independent samples 
with $L=180$ and $m_{0}=0.02$. Since the two species of spins are
symmetry,
the effective sample number is doubled.
Samples of each point are grouped into $4$ runs, and the errors are
obtained from them.  
The parameter $y$
is increasing from the bottom to the top.
The lines are curves of the power law Eq.($\ref{e8}$) with the 
best-fit exponents $\theta_{m}$ to the magnetizations in the first five
time steps as given in Tab.~1. 
We stop at $t=5$ for that gives
the smallest fluctuation. 
Tab.~1 also contains $\theta_{m}$ for
$L=30$, and $60$, which shows that the finite-size effects 
are smaller than the statistical fluctuation.  
By comparing results for various
$m_{0}$ from $0.02$ to $0.08$,  on lattice $L=60$, one can see that
the 
$\theta_{m}$ are quite stable. 
At the decoupled point $y=1$, the Ashkin-Teller model reduces into two
independent Ising models. Our best value $\theta_{m}=0.191(2)$ should 
be compared with the existing numerical results for the Ising model 
from Refs. \cite {oka97a,gra95}, 
$\theta=0.191(1),0.191(3)$, and from those
obtained from auto-correlation before \cite {hus89,hum91}.

\begin{table}[t]\centering 
$$
\begin{array}{|c|c|c|c|} 
\hline
  ~     &       L=180    &      L= 60    &       L= 30    \\ \hline        
  y     &   e_{0}=0.02   &  e_{0}=0.02   &   e_{0}=0.04  \\ \hline
  0     &   -.021(6)     &  -.022(5)     &   -.021(3)     \\ \hline
  1/6   &   -.036(6)     &  -.037(6)     &   -.036(3)     \\ \hline
  1/3   &   -.082(7)     &  -.085(5)     &   -.083(3)     \\ \hline
  1/2   &   -.169(6)     &  -.172(6)     &   -.171(3)     \\ \hline
  2/3   &   -.308(9)     &  -.312(11)    &   -.306(4)     \\ \hline
  5/6   &   -.510(9)     &  -.504(10)    &   -.498(5)     \\ \hline
  1     &   -.812(30)    &  -.793(29)    &   -.760(13)    \\ \hline
\end{array}
$$
\caption{
Results for $\theta_{e}$ with initial condition II
of $e_{0}=0.02$ on $L=60$ and $180$, and of $e_{0}=0.04$ on $L=30$.
The electric order
in the time interval $1\le t \le 5$ is used for the least-square fit
to 
the power law.
}
\label{t2}
\end{table}

Figure~2 shows the initial relaxation of the electric order starting
from 
the initial condition II with $e_{0}=0.02$, $L=180$
in the double-log scale. Each point
is an average over not less than $80000$ independent samples.
To obtain stable results for $y=1$, we used up to $480000$ independent
samples. The lines from the
top to the bottom are curves of the power law Eq.($\ref{e9}$) 
with the best-fit $\theta_{e}$, as given
in Tab.~2, corresponding to $y=0, 1/6$, $...$, $1$,
respectively. For $y\le 1$, the order monotonously
decreases with time in a power law. Discontinuity is found around
$y=1$,
which is the turning point of coupling $K_{4}$ from positive to
negative.
For $y=7/6$, the order jumps to a negative value at the first time step.
The fluctuation for the electric order is much bigger 
than that for the magnetization since the electric order decays 
rapidly to 
very small values. That is the reason why we have to measure
the exponent $\theta_{e}$ at the very beginning of the relaxation
($1\le t \le 5$).

Now, we turn to the measurement of the dynamic exponent $z$. 
Traditionally, $z$ is measured from the long-time exponential decay of
the time correlation or the magnetization of the systems 
\cite {wil85,wan91}.
Due to the critical slowing down, this is somehow difficult. Recently,
a few methods have been proposed to estimate $z$ in the short time
stage
of dynamic processes. Stauffer suggested to obtain $z$ from the power
law
decay of the magnetization with the known static exponent $\beta/\nu$
as
input. For a large enough lattice, one may expect a power law decay of
the orders
\begin{equation}
M(t)\sim t^{-\beta_{m}/\nu z_{m}}
\label{e10}
\end{equation}
\begin{equation}
E(t)\sim t^{-\beta_{e}/\nu z_{e}}
\label{e11}
\end{equation}
before the exponential decay starts. For the Ising model and the Potts
model with ordered initial states, it has been confirmed that the
power
law decay happens in the short-time regime \cite {sta92,mue93,li96,sch96}.
The exponent $z$ may also be measured independently from
the time-dependent Binder cumulants in the short-time regime with
either ordered or disordered initial states as proposed in Refs.
\cite {li95,li96}.
Since higher order moments are needed, the results are relatively more
fluctuating. With the initial dynamic exponent $\theta$ in hand, Ref.
\cite {sch95} suggested to determine $z$ from the power law behaviour of
the autocorrelation. In this paper we do not use this method since the
$\theta_{e}$ for the Ashkin-Teller model is big and difficult 
to be precisely evaluated.

\begin{table}[h]\centering
$$
\begin{array}{|c|c|c|c|c|} 
\hline
~&\multicolumn{2}{|c|}{L=64}&\multicolumn{2}{|c|} {L=90} \\ \hline
y & z_{m}& z_{e} & z_{m}& z_{e} \\ \hline
0&     2.23(4) & 2.24(4)  &  2.21(2) & 2.21(3)\\ \hline
1/6&     2.27(6) & 2.37(6) &   2.24(5) & 2.34(5)\\ \hline
1/3&     2.31(2) & 2.37(2) &   2.29(4) & 2.35(5)\\ \hline
1/2&     2.31(4) & 2.32(4) &   2.26(3) & 2.27(3)\\ \hline     
2/3&     2.26(2) & 2.24(2) &   2.24(4) & 2.23(3)\\ \hline     
5/6&     2.19(2) & 2.17(2) &   2.20(1)&  2.19(1)\\ \hline     
1&     2.17(1)  &2.17(1)  &  2.15(2) & 2.15(2)  \\ \hline   
7/6&     2.12(3)&  2.14(3) &   2.14(3)&  2.16(4)  \\ \hline   
\end{array}
$$
\caption{
Results for $z$ obtained by fitting the power-law 
decay of the order parameters from an ordered initial state with
$L=64$ and
$L=90$, where $z_{m}$ and $z_{e}$ are corresponding to the
magnetization
and the electric order respectively. 
The exact values of $\beta_{m}/\nu$ and $\beta_{e}/
\nu$ are used as input. To avoid the possible deviation from the
power-laws
in the beginning of the relaxation, the first $100$ time steps are
skipped
in the fitting.
}
\label{t3}
\end{table}

In Figs.~3 and 4, we plot the time evolution with the ordered initial
state
for the magnetization and the electric order respectively,
in double-logarithmic scale. Each point is measured 
from $16000$ samples on a lattice with
$L=90$. The curves are the power laws as given by Eqs.
($\ref{e10}$) and ($\ref{e11}$)
with the best fit powers for various $y$. To avoid effects of
$t_{mic}$,
the first $100$ time steps have been skipped. With the static
exponents
$\beta_{m}/\nu$ and $\beta_{e}/\nu$ given by Eq.($\ref{e3}$), 
we obtain
the dynamic exponent $z_{m}$ and $z_{e}$ 
from the powers of Fig.~3 and 4 respectively. The results are compared
with those of $L=64$ in Tab.~3.  The finite-size effects are about
$2.5\%$, comparable with the statistical errors. 
As $y$ varies from zero to $7/6$, a remarkable change
of $z$ may not be accounted only to finite-size effects and
statistical 
errors. This may indicate that the exponent $z$ is varying.
However, within the statistical 
errors, we can not distinguish between $z_{e}$ and $z_{m}$. 
At the decoupling point $y=1$, we obtain the dynamic 
exponent $z=2.15(2)$ from both magnetic and 
electric powers, which  
is consistent with recent result for the Ising model 
obtained by other authors 
\cite {wil85,wan91,sta92,mue93,sil96,bau81}. 

We also estimate the dynamic exponent $z$ independent of the static 
exponents by the finite-size scaling fit of the 
time-dependent cumulants on lattices with $L=64$ and $L=90$. The
method has
been described in detail in Ref. \cite {li96}.
We get the mean value $z=2.16(6)$ from the global fit
and $z=2.16(8)$ from the local fit. The large fluctuation prevents us
to make any conclusion on whether $z$ is varying. Using the mean value
$z=2.16$ as input, the static exponents $\beta_{m}/\nu$ and
$\beta_{e}/\nu$
can be estimated from the finite-size scaling of the magnetization and
the electric order respectively. The results turn out to be consistent
with the analytic results with deviation smaller than $10\%$.

In conclusion, we have confirmed the initial scaling for the kinetic 
Ashkin-Teller
model which is known to possess weak universality in the equilibrium.
The power laws fit to the measurements remarkably well
in the short-time evolution of the order parameters with both
disordered and ordered initial states. The exponents $\theta_{m}$ and
$\theta_{e}$ are found varying with
$y$. The initial order {\it increase} is only 
observed for $y> 1/6$ with the initial condition I.
This fact can not be explained by the straightforward mean-field
argument
that claims the initial order increase should happen when the critical
temperature is lower than the mean-field critical temperature. It
seems
that more careful analysis is needed.

For bigger $y(>1/6)$, $e_{0}$ has negative dimension $x_{e}$.
This means that the time scale associated with $e_{0}$ is a short-time
scale, in contrast with the long-time scale associated with $m_{0}$.
In Fig.~2 one can see that the scaling window gets narrower and
narrower as $y$ increases (the dimension $x_{e}$ decreases). When the
time
scale of $e_{0}$ is comparable with the microscopic time scale,
one will not
see the initial power law behaviour of the electric order.
On the other hand, for $y=0$ 
the electric order and the magnetization have the
same value for $\theta$ within the fluctuations. 
This should be clear since at that point $K=K_{4}$,
i.e., there is symmetry between two orders.

The data in Tab.~3 indicate that the exponent $z_{m}$  is
approximately
equal to $z_{e}$. This may imply that the model has only one time
correlation
length which goes to diverge at the critical temperature, as there is
only
one space correlation length in the equilibrium Ashkin-Teller model as
indicated by the equality of $\nu$ for two orders.

The exponent $z$ may vary with $y$, however, the errors
are still too big to allow a definite conclusion. There are still
some 
interesting questions left for further study. For instances, 
one should better understandthe influence of the choic of 
the initial conditions, 
the discontinuity of the electric order at the decoupling point,
the tricritical point $y=0$, etc.
To locate the critical line segments which 
are not exactly known would be also a good way to test the power of 
short-time dynamics. 

%\bibliographystyle{tex:inputs/misc/pr_np}
%\bibliography{/ising}

\newpage
{\raggedright {\sf Figure captions\\}}
\bigskip
\begin{enumerate}
\item{ 
Magnetization versus time for the initial condition I
with $m_{0}=0.02$, $e_{0}=0$, and $L=180$. The curves from the bottom
to the top
are best fits to the power law Eq. (8) with 
$y=0$, $1/6$, $1/3$, $1/2$, $2/3$,
$5/6$, $1$, and $7/6$, respectively.
}

\item{ 
Electric order versus time for the initial condition II
with $m_{0}=0$, $e_{0}=0.02$, and $L=180$. The curves from the top to
the bottom
are best fits to the power law Eq. (9) with $y=0$, $1/6$, $1/3$, $1/2$, 
$2/3$, $5/6$, and $1$, respectively.
}

\item{ 
The magnetization for various y on $L = 90$, left for relaxation from
the ordered state. The curves are best fits to the power law Eq.~ (10)
with $z_{m}$ given in Tab.~3 and $\beta_{m}/\nu$ given by Eq.~ (3) as
input.
}

\item{ 
The electric order for various y on $L = 90$, left for relaxation from
the ordered state. The curves are best fits to the power law Eq.~ (11)
with $z_{e}$ given in Tab.~3 and $\beta_{e}/\nu$ given by Eq.~ (3) as
input.
}
\end{enumerate}

\newpage

\begin{figure}[p]\centering
\epsfysize=12cm
{{\epsffile{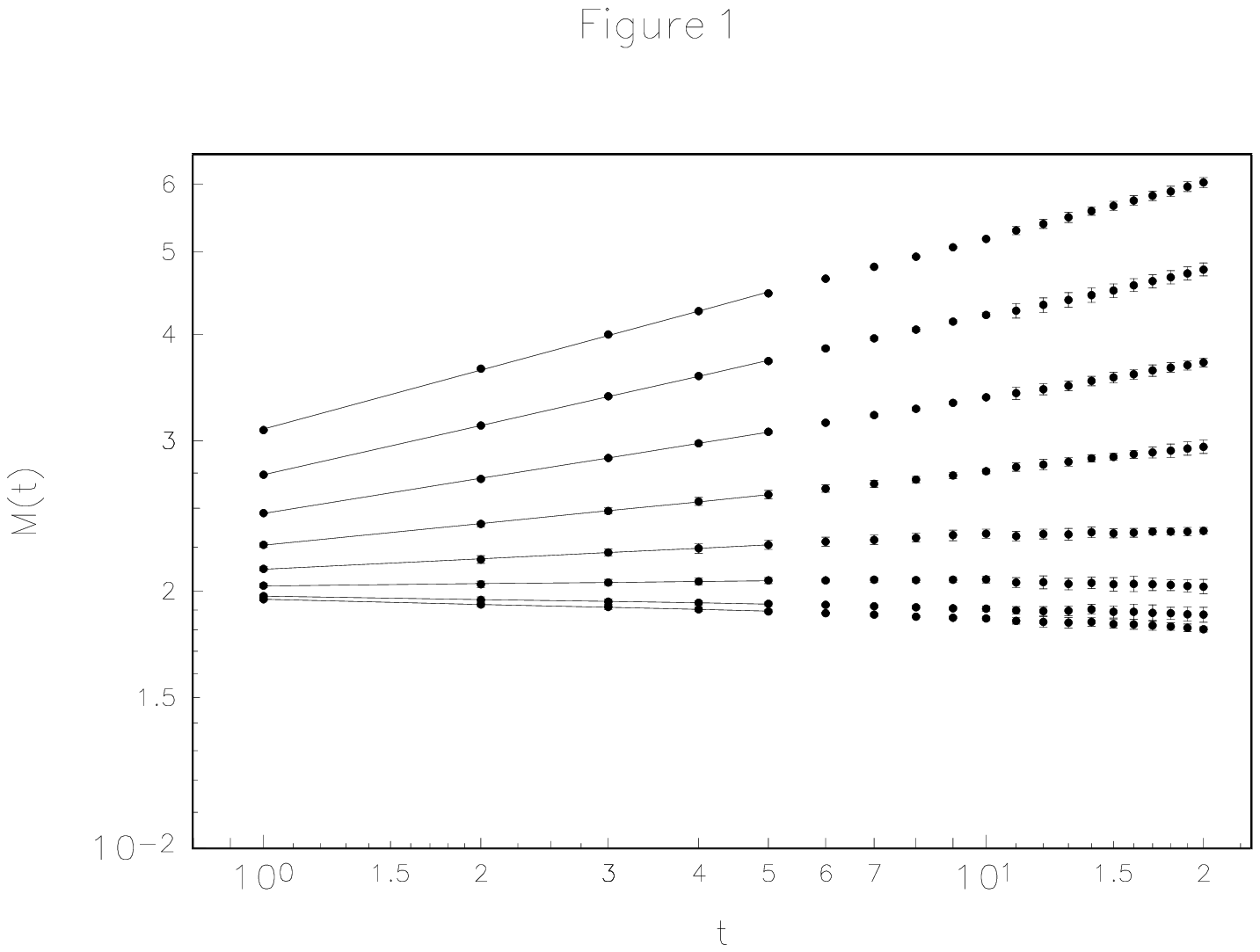}}}
\end{figure}

\begin{figure}[p]\centering
\epsfysize=12cm
{{\epsffile{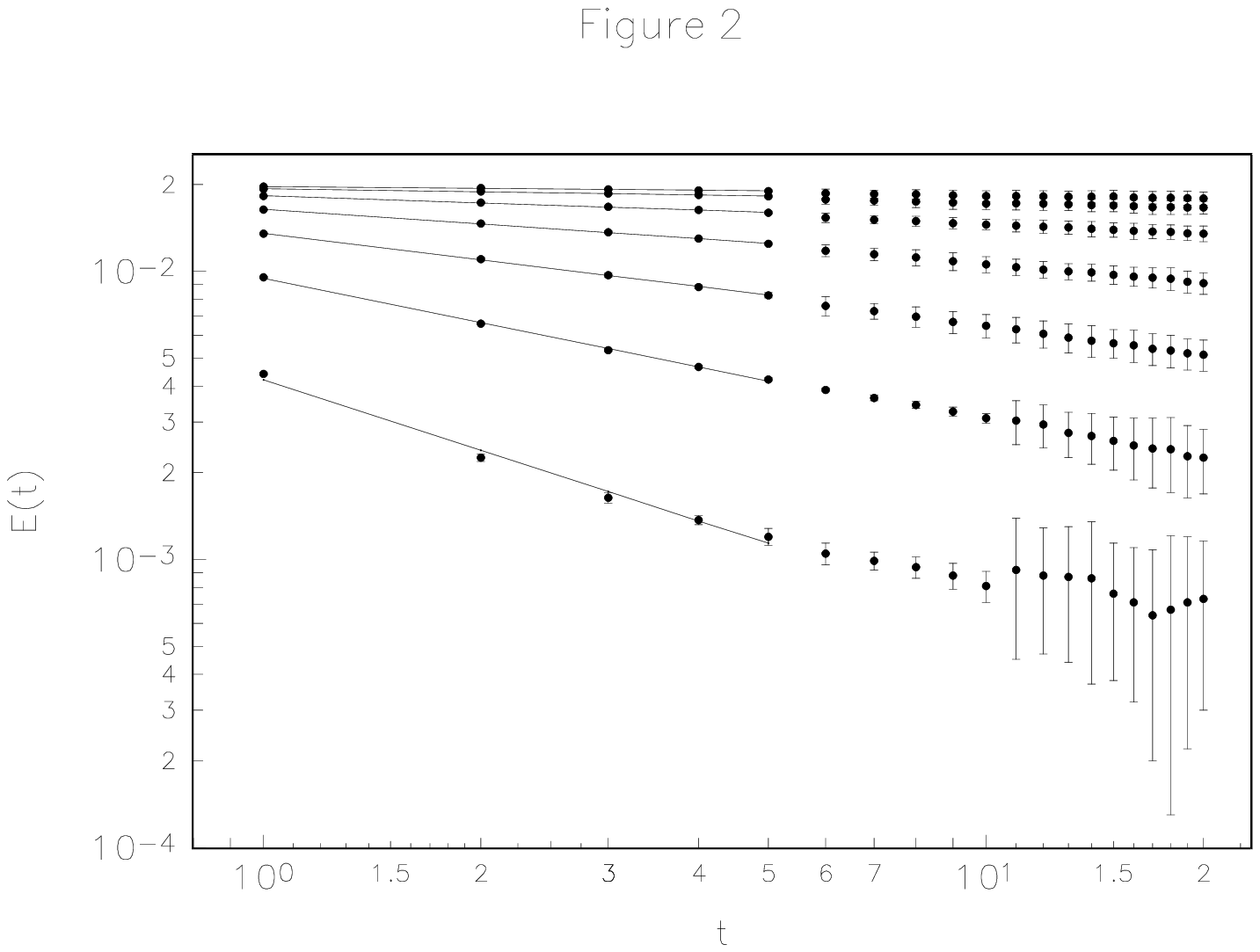}}}
\end{figure}

\begin{figure}[p]\centering
\epsfysize=12cm
{{\epsffile{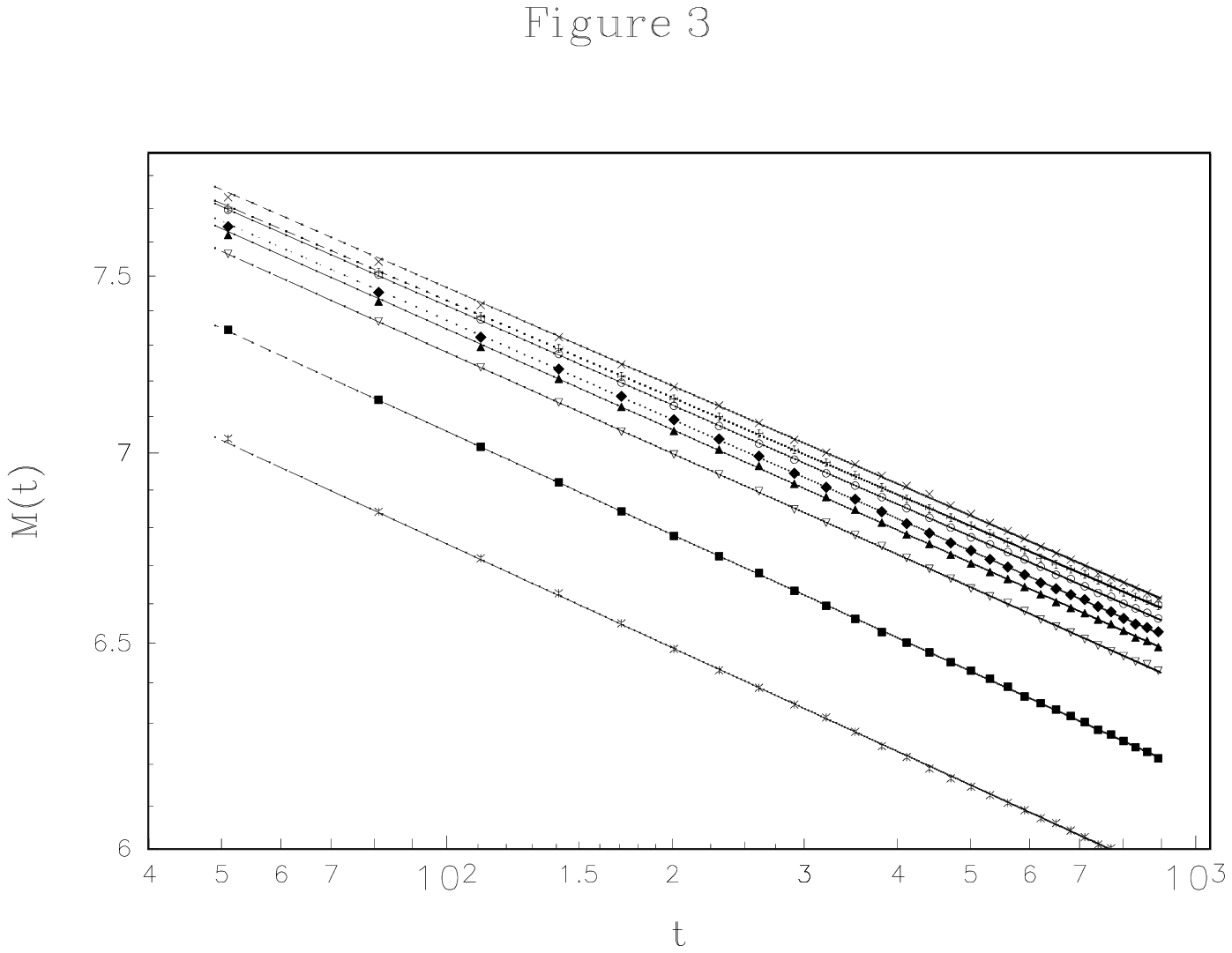}}}
\end{figure}

\begin{figure}[p]\centering
\epsfysize=12cm
{{\epsffile{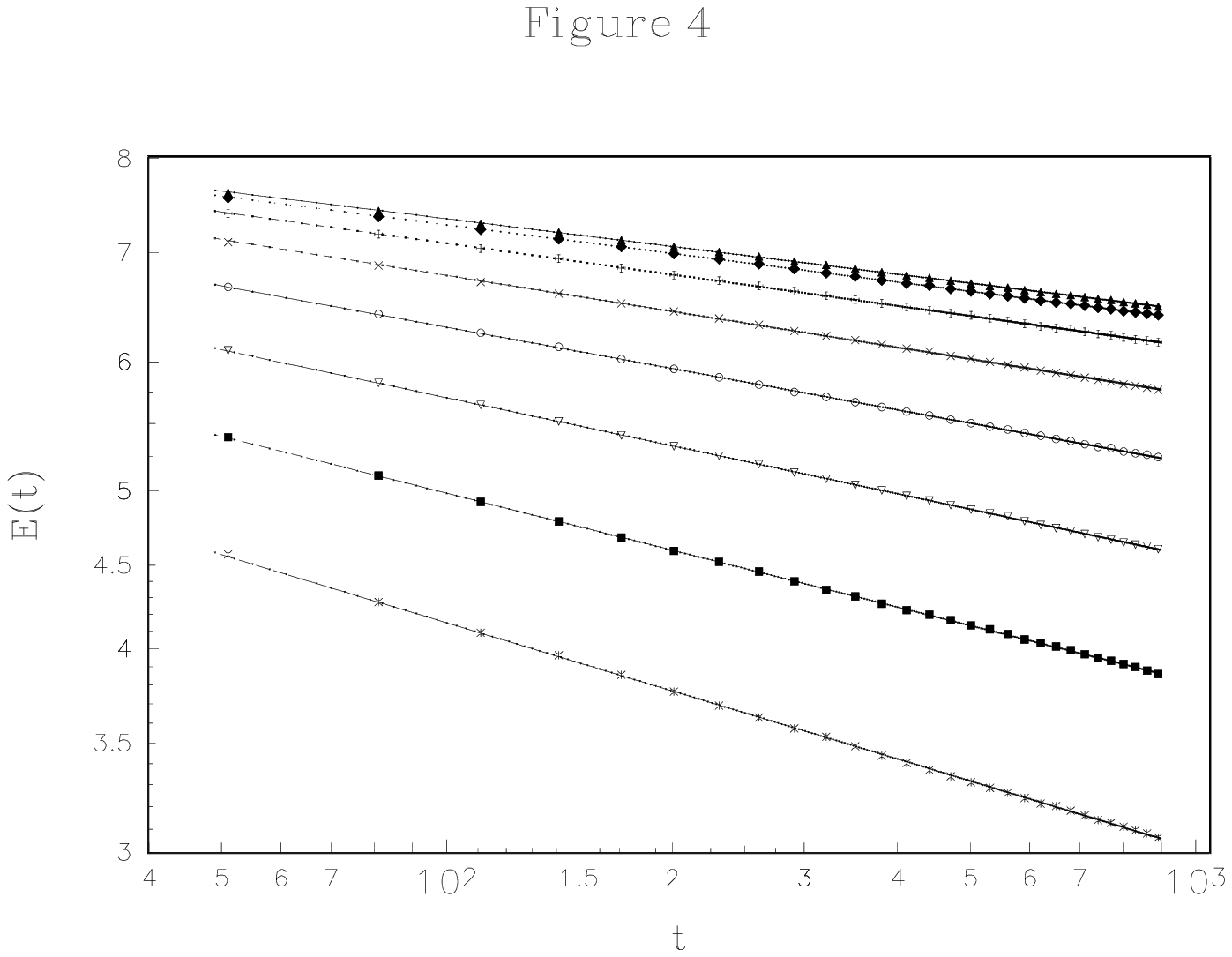}}}
\end{figure}

\end{document}